\documentclass{statsoc}
\usepackage{amssymb,natbib}

\newcommand{\half}{\mbox{$\textstyle \frac{1}{2}$}}

\title[Shapes of Quantum States]{Shapes of Quantum States}

\author[D.~C.~Brody]{Dorje~C. Brody}

\address{Blackett Laboratory, Imperial College, London SW7 2BZ,
UK}

\begin{document}

\begin{abstract}
The shape space of $k$ labelled points on a plane can be
identified with the space of pure quantum states of dimension
$k-2$. Hence, the machinery of quantum mechanics can be applied to
the statistical analysis of planar configurations of points.
Various correspondences between point configurations and quantum
states, such as linear superposition as well as unitary and
stochastic evolution of shapes, are illustrated. In particular, a
complete characterisation of shape eigenstates for an arbitrary
number of points is given in terms of cyclotomic equations.
\end{abstract}


\input{psfig.sty}

\section{Statistical theory of shape}

The idea of a shape-space $\Sigma^k_m$, whose elements are the
shapes of $k$ labelled points in ${\mathbb R}^m$,  at least two
being distinct, was introduced in a statistical context by Kendall
(1984). Here, it is natural to identify shapes differing only by
translations, rotations, and dilations in ${\mathbb R}^m$
(although there are situations of interest, not to be considered
here, in which the scale is also relevant). However, this
identification will not apply to reflections. Thus, the resulting
shape space is the quotient
\[
\Sigma^k_m = S^{m(k-1)-1}/SO(m)
\]
of the sphere by the rotation group. This is the base space of a
fibre bundle with total space $S^{m(k-1)-1}$ and fibre $SO(m)$.
The former is naturally endowed with a uniform Riemannian metric,
while the latter, being a compact Lie group, possesses an
invariant metric.

The key observation of Kendall is that the metrical geometry of
these quotient spaces, long studied by geometers, is precisely the
required tool for the introduction of measures appropriate for the
systematic comparison and classification of various shapes.

Now, for a planar distribution of points, with $m=2$, the shape
space $\Sigma^k_2$ of $k$ points is simply a complex projective
space ${\mathcal P}^{k-2}$ of dimension $k-2$, with the
Fubini-Study metric defining the geodesic distances between pairs
of shapes. The vertices of a shape in ${\mathbb R}^2$ (viewed as a
complex plane), with the centroid at the origin, determine the
homogeneous coordinates of a point in ${\mathcal P}^{k-2}$. The
permutation group $\Sigma(k)$ of order $k$ interchanges
homogeneous coordinates, and is therefore represented by
projective unitary transformations of ${\mathcal P}^{k-2}$.

For three-point configurations, i.e. $k=3$, the shape space is a
complex projective line ${\mathcal P}^1$, which, viewed as a real
manifold, is a two-sphere $S^2$. Hence, the natural shape space
for triangles is essentially a Riemann sphere (Kendall 1985,
Watson 1986).

\section{Space of pure quantum states}

A quantum state $|\psi\rangle$ of a physical system is represented
by a vector in a complex Hilbert space ${\mathcal H}^{n+1}$ of
dimension, say, $n+1$. Quantum states determine expectations of
physical observables, that is, if $X$ is a random variable, then
its expectation in a given state $|\psi\rangle$ is determined, in
the Dirac notation, by
\[
\langle X\rangle = \frac{\langle\psi|X|\psi\rangle}
{\langle\psi|\psi\rangle} ,
\]
where $\langle\psi|$ denotes the complex conjugate of the vector
$|\psi\rangle$. Notice that this expression is invariant under the
complex scale change $|\psi\rangle\to\lambda|\psi\rangle$, where
$\lambda\in {\mathbb C}-\{0\}$. Thus, a pure quantum state is an
equivalence class of states, i.e. a ray through the origin of
${\mathcal H}^{n+1}$. This is just the complex projective space
${\mathcal P}^n$, with the Fubini-Study metric that determines
transition probabilities. To see this, we recall that the
transition probability between a pair of states $|\psi\rangle$ and
$|\eta\rangle$ is given by
\[
\cos^2\half\theta = \frac{\langle\psi|\eta\rangle
\langle\eta|\psi\rangle} {\langle\psi|\psi\rangle
\langle\eta|\eta\rangle}.
\]
To recover the Fubini-Study metric we set $|\eta\rangle =
|\psi\rangle + {\rm d}|\psi\rangle$ and $\theta={\rm d}s$, and
retain terms up to quadratic order (Hughston 1995, Brody $\&$
Hughston 2001). In particular, the maximum separation between a
pair of states is given by $\theta=\pi$, whereas if the two states
are equivalent in ${\mathcal P}^n$, then $\theta=0$.

Therefore, a shape space of planar $k$-point configurations may be
viewed as a pure quantum state space with $k-1$ generic energy
levels. However, this correspondence applies only to planar
configurations, and not those in higher dimensions. Nonetheless,
all this suggests the possibility of applying quantum theoretical
methods to the statistical theory of shapes, or conversely. I
shall briefly outline some relevant ideas which might prove
fruitful.

\section{Triangles as spin-$\frac{1}{2}$ states}

As noted above, the space of planar configurations of labelled
triangles is ${\mathcal P}^1\sim S^2$, or in quantum theory, the
state space of a spin-$\frac{1}{2}$ particle. Specification of the
Hamiltonian determines a pair of energy eigenstates, identifiable
with the poles of the sphere.

As in quantum mechanics where any orthogonal pair of states can be
the energy eigenstates, any pair of orthogonal triangles can form
the poles of $S^2$. A natural symmetrical choice of the
`triangular eigenstates' would be as follows. Let $z$ be a
nontrivial cubic root of unity; this satisfies the cyclotomic
equation $1+z+z^2=0$, and the complex conjugate of $z$ is $z^2$.
Then the vectors
\[
|\!\vartriangle\rangle = \frac{1}{\sqrt{3}}(1,z,z^2) \quad {\rm
and} \quad |\triangledown\rangle = \frac{1}{\sqrt{3}}(z^2,z,1)
\]
could form two eigenstates. Thus, the corresponding shapes are two
origin-centred equilateral triangles, differing only by two-vertex
interchanges. See Fig.~1 (p.~118) in Kendall's Rejoinder (1989),
displaying the triangles on $S^2$ corresponding to this choice of
basis. The shape of an arbitrary triad of points can be expressed
as a complex linear combination $\alpha|\!\vartriangle\rangle +
\beta|\triangledown\rangle$ of these two states such that
$|\alpha|^2+|\beta|^2=1$.

In many statistical problems, the labelling of the vertices has no
significance, and we can work modulo permutations of points. As
regards the representation of the permutation group $\Sigma(3)$ by
transformations of $S^2$, we let the two-sphere be appropriately
subdivided into six regions (`lunes' in Kendall's term) in a
manner topologically equivalent to the subdivision of the surface
of a cube into six faces. The projective unitary transformations
induced by $\Sigma(3)$ then act in a manner that corresponds to
the the permutations of the three oriented cartesian coordinate
axes. The total group of rigid isomorphisms of the cube has 48
elements, i.e. the six permutations of the oriented coordinate
axes, followed by any of the eight possible combinations of the
reflections. In cases where the labelling of points is of no
significance, the relevant state space is thus reduced to the
`spherical blackboard' of Kendall (1984).

\section{Square and pentagon eigenstates}

If $z$ is a nontrivial quartic root of unity, so that the complex
conjugate of $z$ is $z^3$, then the most symmetrical choice for
the square eigenstates consists of the two orthogonal states
\[
|\square_{1}\rangle = \half(1,z,z^2,z^3) \quad {\rm and} \quad
|\square_{2}\rangle = \half(z^3,z^2,z,1),
\]
identifiable in quantum mechanical terms with the spin $\pm1$
eigenstates. The corresponding shapes are two regular
origin-centred squares, differing by vertex interchanges. In this
case, the third eigenstate (for spin-0) is the degenerate square,
i.e. a line, given by
\[
|-\rangle = \half (1,-1,1,-1) ,
\]
and these three states form an orthonormal basis, so that any
configuration of four points is linearly expressible in terms of
these three states.

For pentagons, on the other hand, if $z$ denotes a nontrivial
fifth root of unity, satisfying the cyclotomic equation
$1+z+z^2+z^3+z^4=0$, with the complex conjugate of $z$ given by
$z^4$, then a symmetric basis consists of the four orthogonal
states
\[
\frac{1}{\sqrt{5}}(1,z,z^2,z^3,z^4),\
\frac{1}{\sqrt{5}}(1,z^2,z^4,z,z^3),\
\frac{1}{\sqrt{5}}(1,z^3,z,z^4,z^2),\ {\rm and}\
\frac{1}{\sqrt{5}}(1,z^4,z^3,z^2,z)
\]
of origin-centred regular pentagons. These states are identifiable
with the eigenstates of a spin-$\frac{3}{2}$ particle in quantum
mechanics.

\section{Construction of general eigenshapes}

In general, the projective space ${\mathcal P}^n$ corresponds to
configurations of $n+2$ points. In this case, the cyclotomic
equation
\[
1 + z + z^2 + \cdots + z^{n+1} = 0
\]
indeed determines an orthonormal set of shape eigenstates.
However, as I shall indicate below, if the number $N = n+2$ of
points is not a prime, then the most symmetrical choice of shape
eigenstates, to be referred to as eigenshapes, will be degenerate
in a sense analogous to that of the square eigenstates considered
in the previous section. For example, in the case of six-point
configurations, two of the eigenshapes are regular hexagons, two
of the eigenshapes are equilateral triangles, and the fifth
eigenshape is just a line. In particular, any sextet of points can
be expressed as a linear combination of these five eigenshapes,
and we do not require five hexagonal shapes to express an
arbitrary configuration.

In order to construct general eigenshapes for an arbitrary number
$N$ of points, we let
\[
z = \exp(2 \pi i/N)
\]
denote a nontrivial root of the cyclotomic equation. Then a
generic eigenshape can be expressed in the form
\[
|\omega_k\rangle = \frac{1}{\sqrt{N}} ( z^{a_1}, z^{a_2}, \cdots,
z^{a_N} ) ,
\]
where $\{a_j\}$ ($j=1,2,\ldots,N$) is a set of integers between 0
and $N-1$, not all of which need  be distinct. When they are
distinct for a given $N$, the corresponding eigenshape
$|\omega_k\rangle$ is nondegenerate.

For a general $N$-point configuration, there are $N-1$ eigenshapes
$|\omega_1\rangle, |\omega_2\rangle, \cdots,
|\omega_{N-1}\rangle$. These are given by the table below,
specifying all the values of $\{a_k\}$ for an arbitrary $N$:

\vskip5pt
\begin{tabular}{c|cccccccc}
& $a_1$ & $a_2$ & $a_3$ & $a_4$ & $\cdots$ & $a_{N-2}$ &
$a_{N-1}$ & $a_{N}$ \\
\hline $|\omega_1\rangle$ & 0 & 1 & 2 & 3 & $\cdots$
& $N-3$ & $N-2$ & $N-1$ \\
$|\omega_2\rangle$ & 0 & 2 & 4 & 6 & $\cdots$
& $N-6$ & $N-4$ & $N-2$ \\
$|\omega_3\rangle$ & 0 & 3 & 6 & 9 & $\cdots$
& $N-9$ & $N-6$ & $N-3$ \\
$\vdots$ & $\vdots$ & $\vdots$ & $\vdots$ & $\vdots$ & $\vdots$
& $\vdots$ & $\vdots$ & $\vdots$ \\
$|\omega_{N-2}\rangle$ & 0 & $N-2$ & $N-4$ & $N-6$ & $\cdots$
& $6$ & $4$ & $2$ \\
$|\omega_{N-1}\rangle$ & 0 & $N-1$ & $N-2$ & $N-3$ & $\cdots$
& $3$ & $2$ & $1$ \\
\end{tabular}
\vskip5pt

Note that the numbers $0,1,2,\ldots,N-1$ successively appear in a
cycle. Therefore, for example, if $N=9$, then the value of $a_4$
in the eigenshape $|\omega_3\rangle$, which in this table is given
by $a_4 = 9$, indicates the number following $N-1=8$ in the cycle.
In the present example, this is the number 0 rather than 9, so
that the sequence represented by $a_4$ is $3,6,0,3,6,0,3,6$.

This cyclic property clearly implies that if $N$ is a prime
number, then by following the sequence of numbers along any given
row or column in this table, one never encounters any repetitions.
Thus, the corresponding eigenshapes are all nondegenerate.
Conversely, if $N$ factors, then,  by definition, repetitions will
appear in some rows and columns, and consequently the
corresponding shape becomes degenerate. Therefore, only if the
number of points is a prime can regular polygons be chosen for all
the corresponding eigenshapes.

Although all orthonormal bases are unitarily equivalent, the
choice given here is preferred not only for aesthetic reasons but
also for the practical purpose of systematically generating bases
that are readily visualised. In a quantum mechanical problem, one
begins by specifying the Hamiltonian. The eigenstates of the
Hamiltonian constitute a naturally preferred basis. On the other
hand, in a problem concerning statistical analysis of shape, there
is no {\it a priori} preferred basis unless further conditions are
specified. Consequently, for a given configuration with a large
number of points, it has hitherto been unclear how one can
systematically `decompose' the configuration into a set of simple
orthogonal components. The present scheme offers one possibility
of achieving this task.

\section{Separation and superposition of shapes}

Given an arbitrary pair of $(n+2)$-point configurations in
${\mathbb R}^2$ we can determine the state vectors in the
projective space ${\mathcal P}^n$ that correspond to these two
configurations. Then, the separation of these two shapes (what
Kendall calls the distance between two shapes) is determined by
the transition probability between the two representative elements
of ${\mathcal P}^n$.

The superposition of different shapes is also useful from the
quantum mechanical point of view. In particular, the set of
eigenshapes constructed in the previous section is complete in the
sense that any shape can be expressed uniquely as a linear
superposition of these eigenshapes. For example, superposition of
a pair of orthogonal equilateral triangles, given by
\[
|\rhd\rangle = \cos\half\theta |\!\vartriangle\rangle +
\sin\half\theta\,{\rm e}^{{\rm i}\phi} |\triangledown\rangle ,
\]
will generate all possible shapes associated with three points. In
the special case where $\theta=\frac{1}{2}\pi$, all the collinear
configurations of three points are obtained by varying the phase
variable $\phi\in[0,2\pi)$. Similarly, an arbitrary four-point
shape can be expressed in the form
\[
|\diamondsuit\rangle = \cos\half\theta |-\rangle + \sin\half\theta
\cos\half\eta\,{\rm e}^{{\rm i}\phi} |\square_1\rangle + \sin\half
\theta \sin\half\eta\,{\rm e}^{{\rm i}\psi} |\square_2\rangle.
\]
Thus, the decomposition scheme considered here allows us to
recover simple parametric families of states that represent the
totality of possible point configurations. It should be evident
how these examples can be generalised to shapes with larger number
of points.

\section{Unitary evolution of shapes and geometric phases}

For a given Hamiltonian, the unitary dynamics of a
spin-$\frac{1}{2}$ quantum state corresponds to a rigid rotation
of $S^2$ around the poles specified by the energy eigenstates.
Thus, to study the unitary evolution of triangles, we must
determine the infinitesimal deformations of a normalised triangle
representing the direction orthogonal to that of a rotation about
the origin.

After diagonalisation, the unitary evolution generated by the
Hamiltonian consists in rotating the vertices of the triangle
about the origin, but at generally different angular velocities.
In general, if the angular velocities are commensurable, then the
trajectory is closed in the projective space, but all the vertices
of the triangle are hereby rotated through the same angle about
the origin. This angle is just the geometric phase associated with
the corresponding quantum state.

\section{Quantum entanglement of shapes}

Another interesting question in the present context is whether
quantal notions such as entanglement could be relevant to
statistical shape analysis. The entanglement concept arises when
two or more physical systems are combined. If a system with $m$
energy eigenstates (corresponding to an $(m+1)$-point shape space)
and another with $n$ energy eigenstates (an $(n+1)$-point shape
space) are combined, one obtains the state space of an
$mn$-eigenstate system (an $(mn+1)$-point shape space).

In particular, if the constituent subsystems are disentangled,
then such states form an $(m+n-2)$-dimensional subspace ${\mathcal
P}^{n-1}\times{\mathcal P}^{m-1}$ of the total space ${\mathcal
P}^{nm-1}$. Thus, for example, if we combine a pair of triangles,
we obtain a disentangled pair equivalent to a four-point
configuration, while entangled states represent generic
configurations of five points. More generally, when a pair of
shapes are combined, then a generic shape that corresponds to a
disentangled state will have four points that are collinear. The
possible statistical significance of this situation constitutes an
intriguing open question. Some examples of disentangled and
entangled shape combinations are shown in the figure below.

%%%%%%%%%%%%%%%%%%%%%%%%%%%%%
%%%%%%%%%%%%%%%%%%%%%%%%%%%%%
\begin{figure}
   \centering
    \psfig{file=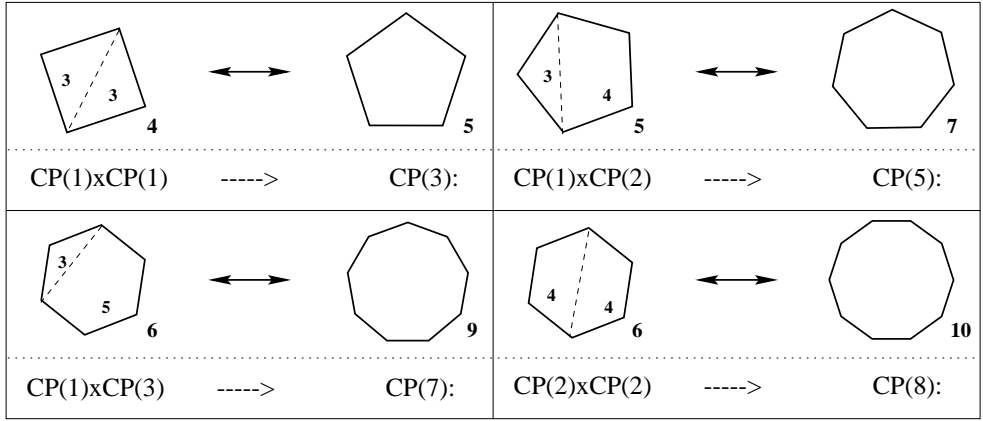,width=13cm,angle=270}
% \makebox{\includegraphics[width=14cm,angle=270]{shape1.eps}}
%
   \caption{Combination of shapes and their entangled
   configurations. When a pair of shapes are combined, the
   resulting shapes of the disentangled states have four points that
   are collinear, as indicated on the left-hand side. The right-hand
   side shapes correspond to entangled states. In general, entangled
   shapes have a larger number of points than the sum of the numbers
   of points associated with individual disentangled shapes.}
\end{figure}
%%%%%%%%%%%%%%%%%%%%%%%%%%%%%
%%%%%%%%%%%%%%%%%%%%%%%%%%%%%

\section{Mixture of shapes}

What would be quite relevant to the statistical analysis of shape
is the notion of mixed states, i.e. distributions over the shape
space ${\mathcal P}^n$. Often, in cases of interest, certain
interactions or dynamics  are associated with the points of the
configurations  under consideration. This permits one to define
distributions of shapes. An example of this is the notion of a
two-dimensional froth, which has important applications in biology
(e.g., epithelial tissue growth). Each froth can be viewed as an
irregular polygon, and the froth vertices where three polygonal
edgesmeet (a vertex with higher incidence number is unstable) are
the points of interest. For a given polygon, the number of edges
can vary, although its expectation value, assuming we have a flat
surface (i.e. zero curvature) and a large number of cells, must be
6 by virtue of Euler's relation (cf. Aste, {\em et al}. 1996).
This is closely related to the fact that, for a wide range of
interaction energies between point particles on a plane, the
minimum energy configuration is typically given by a regular
hexagonal lattice.

In the case of biological cells, for which divisions and
disappearances occur, statistical analysis of the distribution and
the dynamics of point configurations is important in understanding
the properties of biological processes. More specifically, the
statistical theory of shape is relevant here because the ability
of a damaged tissue to restore its stable configuration can be
explained by shape-dependent information stored in the cells, no
further information being required (Dubertre, {\em et al}. 1998).

The problem of shape diffusion (cf. Kendall 1988), which would
characterise the dynamical evolution of froths, can be formulated
as a diffusion process for a single point in ${\mathcal P}^n$. In
the quantum mechanical context this is known as the problem of
quantum state diffusion. One special class of processes that has
been studied extensively (when phrased in the shape-theoretic
context) is as follows: given a set of eigenshapes that represent
energy extremals, (that is, they are eigenstates of an energy
operator), an arbitrary initial configuration will diffuse into
one of the extremal eigenshapes in such a manner that the
probability of terminating in such an eigenshape is given by the
transition probability between the initial and final shapes. For
such a process, an explicit solution to the diffusion equation is
known (Brody $\&$ Hughston 2002) for an arbitrary number of
points. These notions may be applicable to study diffusive
dynamics of complex planar systems.

\section{Discussion}

Finally, I cite some recent developments in the study of point
configurations (Atiyah $\&$ Sutcliffe 2002, Battye, {\em et al}.
2003). These investigations are motivated by the physical question
of determining the minimum classical energy (i.e. most stable)
configurations of point particles in two or three dimensions.
Note, however, that their analysis does not exclude the
possibility of all the points coinciding. Hence, the relevant
shape spaces are slightly distinct from those investigated by
Kendall. These studies, along with the above-mentioned quantum
interpretation of shapes, might shed new light on statistical
shape theory.

\section*{Acknowledgements}

I am grateful to Michael Atiyah for stimulating discussions, and
to The Royal Society for support.

\end{document}